\documentclass[12pt,preprint]{aastex}

\begin{document}

   \title{Massive Star Cluster Populations in Irregulars as Probable Younger
Counterparts of Old Metal-Rich Globular Cluster Populations in Spheroids}

      \author{V.V. Kravtsov}

   \affil{Instituto de Astronom\'ia, Universidad Cat\'olica del Norte,
              Avenida Angamos 0610, Casilla 1280, Antofagasta, Chile\\
              Sternberg Astronomical Institute, University Avenue 13,
              119899 Moscow, Russia\\
              \email{vkravtsov@ucn.cl}
             }

   \begin{abstract}
Peak metallicities of metal--rich (MR) populations of globular
clusters (GCs) belonging to spheroids of different mass fall
within the somewhat conservative $-0.7 \leq$[Fe/H]$\leq-0.3$ range. 
Indeed, if possible age effects are taken into account, this 
metallicity range might become smaller. Irregulars, like the Large 
Magellanic Cloud (LMC),
with longer timescales of their formation and lower star formation
(SF) efficiency do not contain the old MRGCs with [Fe/H] $>-1.0$,
but they are observed to form populations of young/intermediate--age
massive star clusters (MSCs) with masses exceeding $10^4$M$_\odot$.
Their formation is widely believed to be accidental process fully
depending on external factors. From analysis of data available on
the populations and their hosts, including intermediate--age populous
star clusters in the LMC, we find that their most probable
mean metallicities fall within $-0.7 \leq$ [Fe/H] $\leq -0.3$, as
the peak metallicities of MRGCs do, irrespective of sings of
interaction. Moreover,
both the disk giant metallicity distribution function (MDF) in the
LMC and the MDFs for old giants in the halos of massive spheroids
exhibit significant increasing toward [Fe/H]$\approx-0.5$. That is
in agreement with a correlation found between SF activity in
galaxies and their metallicity. The formation of both the old MRGCs in
spheroids and MSC populations in irregulars probably occurs
approximately at the same stage of the host galaxies' chemical
evolution and is related to the essentially increased SF activity
in the hosts around the same metallicity that is achieved very soon
in massive spheroids, later in lower--mass spheroids, and much more
later in irregulars. Changes in the interstellar dust, particularly in
elemental abundances in dust grains and in the mass distribution
function of the grains, may be among the factors regulating star and MSC
formation activity in galaxies. Strong interaction and merger affecting
the MSC formation play presumably additional role, although they can
substantially intensify the internally regulated MSC formation
process. Several implications of our suggestions are briefly
discussed.

  \end{abstract}


\section{Introduction}

Globular clusters observed in both early--type galaxies and
spheroidal subsystems of spiral galaxies (hereafter spheroids)
exhibit, as a rule, bimodality in the color distribution
\citep{gebkiss,kundwhit1,kundwhit2,larsen1}, which primarily
reflects respective distribution in metallicity of the old clusters.
Presently, a consensus is reached that the bulk of spheroids
experienced in their earlier history two major star formation events
which resulted in the formation of two populations of old GCs,
namely metal--poor (MP) and metal--rich (MR) ones. However, it is
not quite clear what physical process(es) is (are) mainly
responsible for these events. Note that the MPGCs are beyond the
scope of the present paper. As for the populations of MRGCs, their
formation is contradictory and disputable problem. On the one hand,
observations have convincingly shown (see, for example, Schweizer
2002; and references therein) merger and/or interaction of gas--rich
spiral galaxies are able to induce powerful starbursts leading to
the formation of massive (globular) star clusters in the mergers, as
was suggested earlier (e.g., Ashman \& Zepf 1992; Zepf \& Ashman
1993). On the other hand, \citet{forbesetal} and \citet{kipatetal}
put constraints on the mergers' contribution to building up the
populations of MRGCs, and argue that in situ formation of these
globulars takes place, particularly due to multiphase collapse (see
Brodie 2002, as well).

No doubt different processes presently taken into account, such as
merging, multiphase collapse, or even capture of GCs through tidal
stripping \citep{cote}, are able to contribute to populations of
MRGCs in spheroids. However, it is not clear whether the given set
is comprehensive and which of these processes (or another one) can
play a leading role. Irrespective of this, converging evidence about
dependence of peak {\it color} of MRGC populations on luminosity
(velocity dispersion) of their parent galaxies
\citep{kundwhit1,forbesforte,larsen1} seems to support in situ
formation of major portion of the MRGCs.

In contrast to spheroids that have formed the bulk of their stars
very soon, within a few Gyr since the Big Bang (e.g.,
McCarthy et al. 2004; Cross et al. 2004) the LMC and other similar irregular
galaxies evolve more slowly with lower SF efficiency. Some of them
contain genuine old MPGCs, with [Fe/H]$<-1.0$, the numder of which
in a galaxy depends, on average, on the galaxy mass. These GCs have
reliably been reveled in the nearest irregulars of the Local Group,
in particular: the LMC is presently known to number up to 15 of such
objects \citep{dutra}, one MPGC is present in the WLM galaxy
\citep{hodge}, and one very probable MPGC is in NGC 6822
\citep{wyder}. A number of the globulars may populate NGC 4449
\citep{gelatt}. Also, \citet{seth} found old GC candidates in
irregular galaxies in the Virgo and Fornax clusters. However, no
MRGCs, with [Fe/H]$>-1.0$, are observed in such galaxies. At the
same time, they are able to form populations of MSCs, i.e.
super--star clusters and populous star clusters. For more
information about these cluster types see, for example,
\citet{billett}. It is often difficult to understand the real causes
responsible for overall increases or bursts of SF activity in the
galaxies, leading to the formation of the MSC populations with
typical mass of individual clusters near or exceeding
$10^4$M$_\odot$. These events are widely believed to occur
accidentally, with no regularity, and to be the consequences of some (strong)
interactions with neighbouring galaxies or due to mergers.

Being at two extreme ends of Hubble sequence, elliptical and
irregular galaxies exhibit very different characteristics, and seem
to be completely unlike to each other in many respects. At the first
glance, apart from the formation of the mentioned MPGCs in spheroids
and irregulars, there are no obvious similarities between subsequent
stages of their formation and evolution. However, in this paper we
argue and draw attention for the first time that the MRGCs in spheroids
and MSCs in irregulars may be counterparts in the sense that they are
tracers of substantially increased SF activity occurring at approximately
the same stage of the host's chemical
evolution. This implies, in particular, that in addition to the
widely accepted processes likely contributing to the formation of
the MRGC and MSC populations, another one(s) may take place and
(primarily) drive the formation of both mentioned kinds of massive
clusters. We address irregular galaxies since they are more "simple"
and chemically more homogeneous in comparison with spirals that have,
as a rule, a notable metallicity gradient across their disks and are
on average more complicated with processes induced by spiral density
waves, bars, etc.

In \S\ref{spheroids} we consider the basic results obtained on the
formation of spheroids, their implications for ages of MRGC
populations, and suggestions about the formation of MSC populations
in irregulars. In \S\ref{irregulars} we summarize and analyze key
observational data obtained to date on the MSC populations in the
LMC and other irregular galaxies. Discussion is contained
in \S\ref{discussion}. Conclusion and final
remarks are in \S\ref{sufirem}.

\section{Formation of Spheroids: Implications for Their Metal--rich Globular Cluster
Populations} \label{spheroids}

Essential progress is presently achieved in understanding of early
spheroids, including timing of the spheroids' formation as a
function of their mass (velocity dispersion).

By proceeding from key observational data available on QSOs and spheroids,
\citet{granato} show the evolution of these objects can be well
understood if one accepts that spheroids of different mass form the
bulk of their stars on different timescales: the more massive
spheroid, the shorter timescale of its formation. Subsequent observations
have further supported this conclusion. Correlations have been found
between velocity dispersion and age \citep{caldwell}, between
[$\alpha
$/Fe] and velocity dispersion, as well as between the
[$\alpha$/Fe] ratios and mean ages \citep{thomas} of early--type
galaxies. Hence one deduces that more massive galaxies had shorter
timescales of their star formation. Indeed, observations of high
redshift objects reveal galaxies with higher mass and SF rate to
form in the earlier Universe
\citep{bercox,bunker,willott,freudling}.

From the above-mentioned and other numerous data (which we omit
here) concerning the formation of spheroids, ages of MRGC
populations as constituents of the spheroids are expected to depend
on mass of the hosts. \citet{thomas} obtained quantitative estimates
for both the epoch of the highest SF rate and duration of SF in
spheroids with different velocity dispersions. According to the
estimates, for example, age difference between these epochs in very
massive spheroids, with velocity dispersion $\sigma\approx$ 300 km
s$^{-1}$, and in spheroids with $\sigma\approx$ 180 km s$^{-1}$ is
of the order of 4 Gyr. In turn, relationship between high SF
rate and formation of massive star clusters is well established and
known from observations of galaxies in the nearby 
Universe. As for star and GC formation in early spheroids, the 
evidence that both MRGCs and metal--rich stars of their hosts formed 
in the same star formation events, with similar ages
and metallicities has been presented by \citet{harris1},
\citet{durrell}, \citet{forbesforte} and \citet{forbes}. MRGC
populations are typically well studied in massive early-type
galaxies with velocity dispersions usually falling in the mentioned range,
i.e. in spheroids populated, as a rule, with large number of MRGCs. Available
direct estimates of age of the MRGC populations in such galaxies
\citep{puzia1,larsen,puzia2,beasley,strader} do not rule out
possible trend of the populations' mean age with mass of the hosts. 
However, strictly speaking, actual accuracy even of spectroscopic 
age estimates is around $\pm$ 3 Gyr at best. For this reason, one 
typically concludes that MRGC populations in galaxies are older 
than 8--10 Gyr. That is not sufficient to reliably judge whether
there is systematic difference of mean age of the populations in
galaxies of different mass. Nevertheless, the systematic difference
in timescale of the formation of early--type galaxies having different
velocity dispersion \citep{thomas,caldwell} allows us to accept, to
a first approximation, the same trend for systematic age variations among 
MRGC populations in the galaxies in the sense: the more massive 
galaxy (or the higher its velocity dispersion) the older on average its 
MRGC population.

Note also that a consensus is presently achieved concerning close
similarity between MRGC populations in early--type galaxies and
their counterparts in spirals: \citet{forbesetal1} demonstrate that
more accurate and reliable present--day investigations have
significantly reduced previously believed discrepancies between peak
metallicities of MRGCs populating both types of galaxies.

Most reliable spectroscopic or/and photometric estimates of peak
metallicities of MRGC populations in well--studied galaxies of the
nearby Universe, such as M 31, Galaxy, NGC 5128, M 81, NGC 4594 are
very close to [Fe/H]$\approx-0.5$ and indistinguishable from each
other within error \citep{barmby,perrett,cote99,woodley,ma,lfb}. On
the other hand, peak color of MRGC populations, in particular
frequently used the ($V-I$) color is observed to correlate with
parent galaxy velocity dispersion (galaxy luminosity) in the sense:
the higher velocity dispersion (luminosity) the redder peak color
\citep{forbesforte,larsen1,kundwhit1}. This point has to be discussed
in some detail. We consider here, for example, latest
results on the dependence of peak color of MRGC populations on luminosity
(stellar mass) of parent galaxies, which appeared while the present paper
was at the reviewing stage. In their impressive work \citet{peng}
have studied the ($g-z$) color distributions of GC systems by relying on
essentially increased number and luminosity range of early--type
galaxies belonging to the Virgo cluster. Using their calibration of
the ($g-z$)--[Fe/H] relation for GCs the authors deduce the peak
metallicity systematic scatter among MRGC populations isolated in
groups of the least and most massive galaxies to be
$\Delta$[Fe/H]$\approx$0.5 dex. Specifically, they estimate the peak
metallicity to vary between $-0.7 \leq$[Fe/H]$\leq-0.2$ (see their
fig. 13) for the groups of galaxies which have mean stellar masses
differing by nearly 2.5 order of magnitude. We note here that the
estimated metallicity range has to be considered as {\it upper}
limit because ages of the MRGC populations have been assumed to be
approximately the same, and the populations' color trend is fully
attributed to their metallicity trend. \citet{peng} mention,
however, that age difference of 10 Gyr between the populations in
the least and most massive galaxies would require if the color trend
was caused solely by age trend.

Finally, mention that \citet{durrell} have arrived at conclusions
about: (i) close similarity between the MDFs in the halos of NGC
5128 and M 31; (ii) a sharp increase of the MDFs somewhere at
[Fe/H]$>-0.8 - -0.7$; (iii) approximately the same metallicity at
which these MDFs of the field stars and the MDFs of the respective
MRGC populations achieve their maxima. Of course, caveats concerning
MDFs of the field stars in galaxies should be taken into account.
Any such a MDF may be affected by numerous factors related, in
particular, to observational uncertainties and to the properties of
the constituent stellar populations, including the spatial
distribution of the populations in the galaxies. The latter factor
is among those that can significantly change an observed MDF and its
features. For instance, the MDFs for the Galactic outer halo and the
Galactic bulge differ dramatically since the bulk of stars in these
Galactic regions belong to metal--poor and metal--rich components of
the MDF, respectively. However, the MDFs of the field stars in large
radial range in a (massive) spheroid, like NGC 5128 \citep{harris3},
exhibit the majority of stars to belong to the metal--rich component
with its maximum occurring around [Fe/H]$\approx-0.5$ ($\sim$
Z$\odot$/3). At the same time, the metal-rich component is broadened
and shows some uncertainty of its maximum's location in metallicity,
including dependence of this location on radial distance in the
galaxy. Despite this, the metal-rich components of the MDFs in the
halo of both M 31 and  NGC 5128 reveal the locations in metallicity
of their maxima approximately corresponding to each other and to the
maxima's locations of the MDFs of the respective MRGC populations in
the galaxies. In contrast to the MDFs of the field stars, the MDFs
of GCs exhibit different radial dependence in galaxies. The peak
positions of both MPGCs and MRGCs in their color distribution have
been found to remain constant with radius in a sample of
ellipticals, despite the change of the relative number of the two
cluster populations \citep{larsen1}.

Taken together, the above results imply that both active phase of the
formation of the MRGC populations and the essentially increased SF
activity in the hosts occurred in the majority of early
spheroids as soon as they achieved approximately the same
stage of their chemical evolution. If so, then the following
question arises: what does (did) occur in other galaxies,
particularly in irregulars passing (passed) the same "particular"
stage of their evolution?

\section{Massive Star Clusters in Irregular Galaxies} \label{irregulars}

\subsection{The intermediate--age populous star clusters in the Large Magellanic Cloud}

If the formation of MRGCs in different speroids is really related to
the same (or approximately the same) stage of their chemical
evolution then the processes, similar to those passed effectively
and rapidly in early spheroids, may be expected to occur later, with
lower efficiency in irregular galaxies because of much longer
timescales and lower efficiency of their star formation. A number of
observations seem to support the suggestion.

First of all we consider the LMC as a well studied irregular galaxy
known to have the present day mean metallicity not less than [Fe/H]
$\approx-0.3$ \citep{luck}, and to contain a large number of
intermediate--age populous star clusters with typical mass of the
order of $10^4$M$_\odot$ or even somewhat higher.

\citet{harris2} have shown (see their Fig. 1, and references
therein) that the MDF for the outer disk stars of the LMC, obtained by
\citet{cole00}, is virtually identical with those for the old red
giant stars in the halo of NGC 5128 and in the halo of M 31. It
reaches its maximum near [Fe/H]$\approx-0.5$, as well. It has to be
noted that despite high accuracy of the data on metallicities obtained from
spectroscopy by \citet{cole00} as compared with similar data based
on photometry, the respective sample of stars studied by them is very
limited (39 giant stars), whereas each sample of stars representing MDFs
in the halos of NGC 5128 and M 31 includes more than 500 giants. However,
a new MDF obtained by \citet{cole05} from spectroscopy of much more
numerous giants (373 stars) situated closer to the galaxy
center, in the LMC bar, exhibits the same behaviour and obvious
maximum around [Fe/H]$\approx-0.4$ that is in good agreement with
data of \citet{cole00}. A slightly more metal--rich value of the MDF
maximum's location in metallicity is fairly clear, taking into
account more central position of stars to the galaxy center, as well
as our above discussion on the MDFs in spheroids.

According to recent conclusions of \citet{geisler}, a mean
metallicity of the populous star clusters formed in the LMC 1--3 Gyr
ago is close to [Fe/H]=$-0.5$, {\it irrespective of their age} (see
also Fig. 3 in review by Da Costa 2002). That is, it was not
changing during the period of the cluster formation. At the same
time, data on metallicities and ages determined on a homogeneous
scale from color-madnitude diagrams of both the intermediate--age
(age $>$ 1 Gyr) clusters and intermediate--age field stars near
these clusters \citep{piatti} reveal surprising and important
difference between the age--metallicity relations of the clusters
and field stars. We demonstrate this difference in Fig.~\ref{agemet}
based on data of \citet{piatti} on the clusters and star fields
(filled squires), as well as on data of \citet{olszewski} on
additional sample of the clusters (asterisks). While the cluster
metallicities do not exhibit obvious dependence on age (lower
panel), as shown by \citet{geisler}, the field star metallicities do
it (upper panel; note that five star fields with the same
metallicity of [Fe/H]=$-0.25$ and age of 1.3 Gyr are shown by one
square in the panel). Additionally, the data (from table 3 of Piatti
et al. 2003) demonstrate that the MDF for the given sample of the
LMC populous star clusters is very similar to the typical MDFs of
MRGSs in spheroids, and that the majority of the clusters fall
within $-0.7 \leq$ [Fe/H] $\leq -0.3$.

This implies that in contrast to the field stars, the most massive
intermediate--age star clusters (i.e., star clusters which survived
more than 1 Gyr since their formation) in the LMC  did not "feel"
systematic enrichment with metals of the interstellar medium in the galaxy
during the epoch when the clusters were forming. In other
words, these clusters formed preferentially from the interstellar
matter with small difference of its metal content, irrespective of
their age, location in the galaxy, as well as possible interactions
of the LMC with its neighbours. Surprisingly, MRGCs in M 31 seem to
exhibit similar effect. From spectroscopic investigation of globular
clusters in M 31 \citet{puzia3} find peak metallicities of
intermediate--age ($5 - 8$ Gyr) and very young ($<$ 1 Gyr) globulars
in the galaxy to be around [Fe/H]=$-0.6$ and [Fe/H]=$-0.4$,
respectively. Remind that the peak metallicity of the old MRGCs in M 31
falls somewhere between $-0.6 \leq$ [Fe/H] $\leq-0.5$
\citep{barmby,perrett}. This suggests that during period of $\sim$12
Gyr or so, the MRGCs in M 31 were forming with approximately the
same peak metallicity, within $-0.6 \leq$ [Fe/H] $\leq-0.4$,
irrespective of age and location in the galaxy.

The LMC is surely the most convincing example in comparison with
other irregular galaxies because there are relatively reliable data
available on the basic characteristics of its star clusters and
field stars. However, the presented important evidence concerning
the formation of the LMC intermediate--age populous star clusters is
not, strictly speaking, sufficient to extrapolate it on MSC
populations in other similar irregulars. For this reason, we have
undertaken a search in the literature for data on the irregular galaxies 
known to host MSC populations.

\subsection{Young massive star cluster populations in irregular galaxies}

To compile our list of the irregulars, we selected them from (i)
list by \citet{larsricht2} who searched for young MSCs in nearby
non--interacting galaxies \citep{larsricht1}; (ii) list by
\citet{billett} who searched for compact star clusters in a sample
of nearby irregular galaxies some of which were found to host MSCs,
irrespective of signs of their interaction; (iii) a number of other
publications devoted to study of MSCs in irregular galaxies. Our
final list consists of twelve galaxies, including the LMC. The
majority of them are the famous examples of starburst galaxies or
galaxies with high SF rate, which form populations of MSCs. The list
is presented in column 1 of Table~\ref{tbl-1}. There are also listed
the galaxy absolute $B-$magnitudes (column 2); the oxygen abundance
(12+log(O/H)) of HII regions in the galaxies (column 3); references
to sources of the data on the oxygen abundance: these are papers
containing either the original data or references to the original
data (column 4); cluster metallicities (column 5); references to
papers in which the MSCs were studied and their metallicities were
either accepted to correspond to the respective hosts' metallicities
or estimated from the best model fit to cluster colors, or were
estimated from color--magnitude diagrams (column 6). In column 7 we
also roughly denote richness of the young/intermediate--age MSC
populations, that is conditionally divided by three grades according
to the number of the clusters with masses around or exceeding
$10^4$M$_\odot$: {\it poor} (10 or less number of the clusters),
{\it medium } (several tens of the clusters) and {\it rich} (around
100 or even more of the clusters). Data are from \citet{larsricht2},
\citet{billett}, and papers referred to in column 6.

It is seen that the oxygen abundances available for the galaxies of
our list (excluding, of course, M 82 as one of the most complicated
cases among starburst irregulars; see lower)  are surprisingly
similar and fall in narrow range, $\Delta$(log(O/H))$<0.25$ dex,
with their formal mean value near 12+log(O/H)=8.28. Very close
similarity of oxygen abundances of the galaxies NGC 4214, NGC 4449,
NGC 5253 were mentioned earlier by \citet{martin}, too. The mean
oxygen abundance as well as individual ones of the galaxies of our
list are surely sub--solar. It has to be noted that
correctly determining the nebular oxygen abundance, especially in
metal--rich environments is a complicated task (see, e.g., Castellanos et
al. 2002; Pilyugin et al. 2003; Pilyugin et al. 2004; and references
therein) because of the problem with deriving electron temperature and
therefore with direct determination of the abundance. In this case,
indirect methods can be used, some of which systematically
overestimate the oxygen abundance. Moreover, somewhat different
quantities for the solar value of 12+log(O/H) are deduced by different
authors. For definiteness, we refer to \citet{leeskill}. According to
them, the mean oxygen abundance of NGC 1705 is 12+log(O/H)=8.21 $\pm 0.05$
and corresponds to [O/H]=$-0.45$ (i.e., the solar value is accepted to be
12+log(O/H)=8.66). In any case it is clear that the mean oxygen abundance
of the galaxies of our list is at least somewhat lower as compared
to the LMC abundance of 12+log(O/H)=8.35 and it falls between
approximately $-0.65 \leq$ [O/H] $\leq -0.35$ in dependence on the
accepted solar value of 12+log(O/H). This approximately corresponds
to 0.2Z$_\odot\leq$ Z $\leq$ 0.4Z$_\odot$ that is nearly identical
to the range between two metallicity values, Z=0.004 and Z=0.008
(with Z$_\odot$=0.02), which are often used in cluster evolutionary
models.

Metallicities of the cluster populations in the galaxies listed in
Table~\ref{tbl-1} have been estimated from the best model fit to
cluster colors. Two exceptions are metallicities
of MSC populations in NGC 1140 and NGC 5253. Due to young age of the
populations their metallicities were assumed to be close to the
host galaxies' actual ones. For this reason the respective metallicity
values are indicated in brackets.
In NGC 1569, metallicity estimates have been made by two
teams using different models. Results of \citet{anders} show that
the majority of the NGC 1569 star clusters with age more than 8 Myr
have metallicities in the range $-0.7 \leq$ [Fe/H] $\leq -0.4$. Star
clusters with age less than 8 Myr are more metal--rich, but, at the
same time, they are systematically less massive. This means that
members of this subpopulation have systematically lower chances to
survive. Two of three most massive star clusters in NGC 1569, with
mass $> 10^5$M$_\odot$, were estimated to have [Fe/H]$\approx-0.4$
([Fe/H]$\approx-1.7$ is deduced for third one). In their study of
the MSCs in NGC 1569 \citet{hunter1} noted "that the Z=0.004 cluster
evolutionary tracks did not account very well for the colors of some
of the clusters, but that the Z=0.008 models did", whereas the
oxygen abundance of emission nebulae in the galaxy implied a
metallicity Z$\approx$0.004 for the star clusters.

Similarly, \citet{gelatt} compared their photometry of the star
clusters in NGC 4449 to cluster evolutionary models and used both
Z=0.004 and Z=0.008. They found that while the Z=0.004 models are
closer to metallicity of NGC 4449, the Z=0.008 models fit cluster
colors better. Also, for the IC 10 star clusters \citet{hunter2}
considers Z=0.004 and Z=0.008 as the most appropriate values of
metallicity, but she concludes "that Z=0.008 clusters evolutionary
tracks better match the observed integrated colors of star
clusters".

There are very limited data on the galaxy NGC 6745 compared to
other galaxies from our list. We failed to find any information
about estimates of the oxigen abundance in NGC 6745. Its Hubble type
is not known exactly. Probably, NGC 6745 is not a Magellanic--type
galaxy. It is not excluded that it is a (late?) spiral \citep{zhu}.
\citet{karachentsev} describe this galaxy as "a peculiar object,
with knots that form a ring--shaped structure embedded in a diffuse
elongated envelope". We have included it in our list since it forms
young MSC population that has recently been studied by
\citet{degrijs1}. The metallicity distribution obtained for the
cluster population shows that the majority of the MSCs have
sub--solar metallicity and that its main peak is at 0.005 $<$ Z $<$
0.01. It is worth of noting
that in the same paper the authors study MSC populations in two
different regions of the galaxy NGC 3310, in its circumnuclear
region and in the main galactic disk. It turns out that the
metallicity distributions obtained for both populations are
statistically indistinguishable. The majority of the star clusters
have Z$<$0.015.

We have also included M 82 in our list since it is among the well
known and most impressive examples of starburst irregular galaxies,
with a large number of (relatively young) MSCs formed within the
last 1.5 Gyr or so. At the same time, the most probable metallicity
of these clusters is a dark matter. The galaxy is very dusty and has
a significant, highly variable extinction. This makes significant
difficulties in obtaining reliable data on its cluster population.
From data on multicolor photometry of MSCs in the so--called M 82 B
star cluster system, \citet{parmentier} were not able to easily
distinguish among cluster metallicities ranging from Z=Z$_\odot$ to
Z=0.2Z$_\odot$. Hence the question is open about the most probable
value of metallicity of the M 82 MSCs. Their generic metallicity is
{\it a priori} accepted, as a rule, to be equal to the solar one
(Z=Z$_\odot$). However, \citet{origlia} obtained detailed stellar
abundances in the nuclear region of M 82 and found that gas and
stars trace very similar iron abundances, the average value being
[Fe/H]$\simeq-0.35$ dex.

The data presented and analyzed here demonstrate close similarity
of the most probable mean metallicities of the MSC populations in
starburst irregular galaxies or in galaxies with substantially
increased SF activity, irrespective of presence or absence of some
signs of interactions. This circumstance becomes more important if
we pay attention to the close similarity between the most probable
mean metallicities of the MSCs in the irregulars and those of the
old MRGCs in spheroids. At the same time, it has to be noted that
the presently available data on the mean (peak) metallicities of the MSCs
in irregulars allow no study of any trend (if any) of the
metallicities with the characteristics of host galaxies.

\section{Discussion}
\label{discussion}

Our suggestion regarding the formation of MRGC populations in
spheroids and MSC populations in irregulars in consequence of (or coinciding with)
substantially increasing SF activity in the hosts while they achieve
the same stage of their chemical evolution implies increasing SF
activity in galaxies with increasing their metallicity.

If so, are there any other independent grounds to say about the
ubiquitous and naturally increasing SF activity in galaxies with
increasing their metallicity? Although these grounds are very
limited, they are already available. In particular, (i) it is a
correlation between SF rate in blue compact dwarf galaxies and their
metallicities \citep{hopkins,kong} that is, in our opinion, one of
supplementary important evidence; (ii) another one may be a
clear correlation found by \citet{zoran} between the extinction of
galaxies and their SF intensity (rate) for various epoches in
redshift range $0.4 \leq$ z $\leq 6.5$, that is in agreement with
increasing metallicity of the interstellar matter with time.

Taking into account both the most probable formation of MSCs and the
essentially increased SF activity in galaxies around
[Fe/H]$\approx-0.5$ may help to explain or to better understand a
number of important observations. Among them are the formation of
MSCs in the disks of isolated spiral galaxies and their preferential
location in the outer parts of the disks \citep{larsricht1};
evidence (contradicting to the predictions of semi-analytic
simulations) for a mass--dependent luminosity evolution at
intermediate redshift among Sa--Sdm/Im galaxies \citep{bohm},
meaning that lower--mass spirals achieve their maximum SF activity
later than higher--mass ones; very different impact on the induced
SF activity or on the formation of massive (globular) star clusters
in merging (merged) or interacting galaxies \citep{read,
boselli,pierce}; starburst phenomenon in isolated galaxies or
galaxies without any sign of interaction. Moreover, we assume that
causal relationship may be between the supposed ubiquitous
essentially increased (or bursting) SF activity in galaxies at Z
$\approx$ Z$\odot$/3 and the same metallicity of the intracluster
gas in galaxy clusters.

We note again that our suggestions  do not concern the much lower
metallicity at which MPGCs formed. They have a number
of dissimilarities as compared to MRGCs.  The populations of MPGCs
and MRGCs are discriminated not only by their kinematic, spatial
distribution, age, etc. Other important distinctions between the two
populations of GCs are revealed, in particular the difference between
star--versus--cluster formation processes occurred at low and high
metallicity. Specifically, the globular cluster--to--field star
formation efficiency at [Fe/H]$<-1.0$ is much higher as compared with
that in metal--rich range, at [Fe/H]$>-1.0$ \citep{harris1,durrell,forte}.
Also, the ratio of the number of MRGCs to the number of MPGCs in
galaxies systematically increases with increasing galaxy mass (e.g.,
Peng et al. 2006). These details suggest, among others, essential dependence
of cluster--to--field star formation efficiency in the metal--rich range
on galaxy mass.
At the same time, we note possible important difference between star
and massive cluster formation modes in the metal--rich range, namely
the demonstrated (in Fig.~\ref{agemet}) probable difference in the
age--metallicity relation for field stars and populous star clusters
in the LMC.
The same difference may have taken place in spheroids, as well. This is
indirectly supported by the discussed (in Sect.~\ref{spheroids}) 
radial dependence of peak color of the metal--rich field stars and apparent 
absence of such dependence for the MRGCs' peak color. If this
dissimilarity is real, then one can suggest its two possible interesting
consequences. Specifically, it should directly naturally result in the
well--known difference between concentration of the MRGCs and field stars
to the spheroids' centers, in the sense that the field stars are more
concentrated than the MRGCs. Indeed, if the peak metallicity of MRGCs in a
galaxy does not change during the period of their formation then at certain
critical mean metallicity of the enriching and contracting gas the MRGCs
stop their formation, whereas the field stars yet continue to form closer and
closer to the galaxy center.
Hence the different concentration of star and GCs to the centers of
ellipticals may take place even in the cases of no merger events and
negligible (or no disruption) of GCs in the galaxies' central parts.
Moreover, this picture implies that field metal-rich stars of the spheroids may,
on average, be systematically somewhat younger and more metal--rich than MRGCs.

We also note that within the framework of our suggestions the Small
Magellanic Cloud (SMC) with its actual mean metallicity consisting
approximately [Fe/H]$\approx-0.7 - -0.6$ dex \citep{dacosta}, is
expected to be entering into the discussed active stage of
galaxy evolution. Indeed, in conformity with conclusions achieved,
for example, by the latter author by relying on data obtained to
date on the SMC intermediate--age star clusters with well--derived
parameters, "a relatively abrupt increase in the cluster abundances
up to approximately the present--day abundance of the SMC field
stars" has probably commenced near 2--3 Gyr ago. \citet{piatti05}
have increased the sample of intermediate--age clusters in the SMC
with well--derived parameters. The available data show that along
with intermediate--age metal--poor ([Fe/H] $<-1.0$) clusters, the
SMC has already formed some amount of more metal--rich ones, at
least some of which are presumably fairly massive star clusters
(taken into account their intermediate age exceeding 1--1.5 Gyr),
with metallicity exceeding [Fe/H]=$-1.0$ and achieving
[Fe/H]$\approx-0.6$. However, their number in the galaxy is probably
still small, as compared to the number of the metal--poor
counterparts. The distribution on metallicity of the SMC
intermediate--age star cluster population seems to resemble the
bimodal distribution on metallicity of GC populations in galaxies,
with a minimum of clusters around [Fe/H]=$-1.0$. Indeed, almost no
clusters somewhere between $-1.1 <$ [Fe/H] $< -0.8$ are observed in
the SMC (for details on the SMC star clusters and the respective
references see, for exemple, review by Da Costa 2002). The
ages--metallicity relation for the SMC intermediate--age star
clusters implies that around 2 Gyr ago, when its metallicity
increased abruptly the SMC apparently finished to form (massive)
clusters which may be ascribed to metal--poor population, and
started to form clusters which, in turn, may be ascribed to
metal--rich one, with [Fe/H] $>-1.0$. To avoid confusion, we note
that it is the latter population that we consider to be counterpart
of the LMC intermediate--age star clusters. Although SF histories in
the SMC and LMC exhibit specific features, they demonstrate
important similarities, too. Specifically, after a long quiescent
epoch of many Gyr the LMC has experienced considerable rise of its
SF activity within the last $3-4$ Gyr when it formed the large
number of its stars, and its metallicity changed from
[Fe/H]$\approx-0.7$ dex to [Fe/H]$\approx-0.3 - -0.2$ dex.
Similarly, possible entering of the SMC into an active phase of its
evolution may be implied by the results of \citet{harzar} on SF
activity of the SMC during its life: a long quiescent epoch between
3 and 8.4 Gyr ago in the galaxy was followed by at least three peaks
in SF rate, at $2-3$ Gyr, 400 Myr, and 60 Myr ago.

Our apparent avoiding various deviating cases regarding GC (MSC)
populations in galaxies does not mean idealization of real
situation. No doubt the real picture is surely complicated due to
the existence of various peculiar cases suggesting that a number of
processes may impact on the formation (or surviving) of GCs (MSCs)
and their systems in galaxies. Observations reveal (or suspect), for
example, a near--complete lack of the MPGCs in IC 4051
\citep{wooworth}, a giant elliptical galaxy in the Coma cluster or
the absence of the MRGC population in the galaxy NGC 4478
\citep{kipatetal02} belonging to the Virgo cluster. In the
well--known case of the dwarf elliptical galaxy M 32 the situation
is even more surprising: no GCs are observed in the galaxy. As for
the MRGCs, it has to be noted that results of \citet{peng} show
systematically very scanty or no MRGC populations in the least
massive early--type galaxies. All these and other possible cases
are, of course, valuable for our comprehensive understanding of the
formation and evolution of MRGCs (MSCs) and their systems.

\section{Conclusion and Final Remarks} \label{sufirem}

This paper is devoted to the formation of old metal--rich GC
populations observed in the majority of spheroids, as well as of the
populations of young/intermediate--age massive star clusters, with
masses exceeding $10^4$M$_\odot$, forming/formed in the Large
Magellanic Cloud and in other irregulars.
Our main goal is to demonstrate that presently available data on the
star clusters and their hosts imply the existence of a factor,
defined by us as "chemical" factor, which may favor and even be
mainly responsible for the formation of both kinds of massive star
clusters and for the accompanied substantially increased SF activity
in the hosts.

In order to interpret and reconcile all the presented and analyzed
observational data on the star clusters and their hosts we conclude
the following.

\noindent {\bf --} {The active phase of the formation of both kinds
of the cluster
populations, MSCs in irregulars and MRGCs in spheroids, probably
occurs at and is related to the same "particular" stage of the host
galaxies' chemical evolution, taking place somewhere between $-0.7
\leq$ [Fe/H]$\leq-0.3$ and occurring in the majority of
galaxies; it is reached very soon in massive spheroids, later in
less massive spheroids, and much more later in irregulars. The
particular abundance of the inter--stellar matter may be a factor
affecting SF efficiency and favoring the formation of massive star
clusters, taking into account the important observation that mean 
metallilicity of the LMC intermediate--age populous star clusters
was not notably changing over a period of the formation of the clusters,
during approximately 2 Gyr. This implies that at least in metal--rich 
range, at [Fe/H]$>-1.0$, the variation of star and MSC formation activity 
is primarily an internally regulated and metallicity--dependent processes. 
In this connection, it is worth of noting two important conclusions made on 
the nature of starburst galaxies in the nearby Universe. Namely, a large
number of the galaxies turn out to be either isolated objects or
those without any direct sign of interaction \citep{coziol}, and the
starburst phenomenon being rather self-sustained phenomenon with
some mechanism of internal regulation \citep{cozioletal}.}

\noindent {\bf --} {We do not exclude that the essentially increased
SF activity and the preferential formation of massive (globular)
star clusters around [Fe/H]$\approx-0.5$ may be due to a favorable
combination of a number of factors, realization of which is most
probable just at the particular chemical composition of interstellar
medium. It is likely that changes in the interstellar dust are at 
least partially responsible for the
increasing SF activity and for the formation of massive (globular) star 
clusters in galaxies with increasing metallicity. Indeed, along with
correlations (mentioned above) between SF activity in galaxies and
their metallicity (extinction), a correlation between the
dust-to-gas ratio and metallicity in the interstellar medium of
dwarf irregular galaxies has been found \citep{schmidt}. At the same
time, the mass (size) distribution of dust particles depends on the 
environment (density) and the gas-to-dust ratio (Kim \& Martin 1995, 
and references therein). Moreover, proportion of dust grains differing
by their properties including elemental abundance 
(see, for example, Chiar \& Tielens 2006) probably vary with the chemical 
evolution. Factor(s) responsible for subsequent decreasing SF
activity in spheroids is (are) not quite obvious and may be somewhat
different in galaxies of different mass. Whether mass loss and/or gas
exhaustion are the only responsible for this decreasing is not clear.
Also, it is now difficult to say why there is a trend of MRGC 
population peak metallicity with parent galaxy mass.}

\noindent {\bf --} {Other important factors can
substantially affect and modulate the cluster formation process.
Along with strong interaction and merger which intensify this process,
a parent galaxy mass also affects it, and low mass of a galaxy being
able to put the limit for the ability of the galaxy to form MSCs
\citep{billett}.
Moreover, the analyzed data hint that mergers and interactions have
their highest efficiency to induce the formation of massive star
clusters just in this limited "favorable" range of metallicity
marked by metallicities of the old metal--rich globular clusters in
spheroids and young (intermediate--age) massive star clusters in
irregulars. We suggest that interacting and merging galaxies convert the
gas with metallicity within the mentioned range to massive star
clusters more effectively than isolated galaxies.

Finally, by proceeding from our analysis and conclusions we expect
that more accurate and detailed data (with the age--metallicity
degeneration disentangled) on metallicities of the known or/and
newly found populations of young/intermediate--age {\it massive}
star clusters in irregulars (including M 82) or spirals, as well as
of intermediate--age globular clusters in early--type galaxies will
show a limited scatter of their most probable mean (peak)
metallicity around [Fe/H]$\approx-0.5$. This would reinforce the
grounds to conclude that in the metal
rich range, at [Fe/H]$>-1.0$, a clue to the formation of (very)
massive star clusters may be rather at "micro--level", i.e. in
internally regulated (externally intensified) and
metallicity--dependent processes, particularly in the interstellar matter,
namely in molecular clouds forming the (very) massive star
clusters. In this connection, it is worth of noting that the same
physical reasons may be among factors leading to the formation of
ultra--compact dwarf galaxies, such as their metal--rich
subpopulation observed in the Fornax cluster, for which
\citet{mieske} have found from spectroscopy the mean metallicity of
[Fe/H]=$-0.62\pm 0.05$ dex. Within error this value virtually
identical with the most probable peak metallicity of MRGCs. Moreover,
most massive MRGCs achieve (or overlap) the lower boundary of
luminosity of the ultra--compact dwarfs.

\acknowledgements We are grateful to an anonymous referee for useful
comments and suggestions resulted in a number of important improvements
in the manuscript. We acknowledge Thomas Puzia for helpful discussion of
basic points considered in the paper. Thanks are due to Marcus
Albrecht for his kind assistance with providing the author with some
sources of the data used in the present paper.

\clearpage

\begin{figure}
  \centering
  \includegraphics[angle=-90,width=8cm]{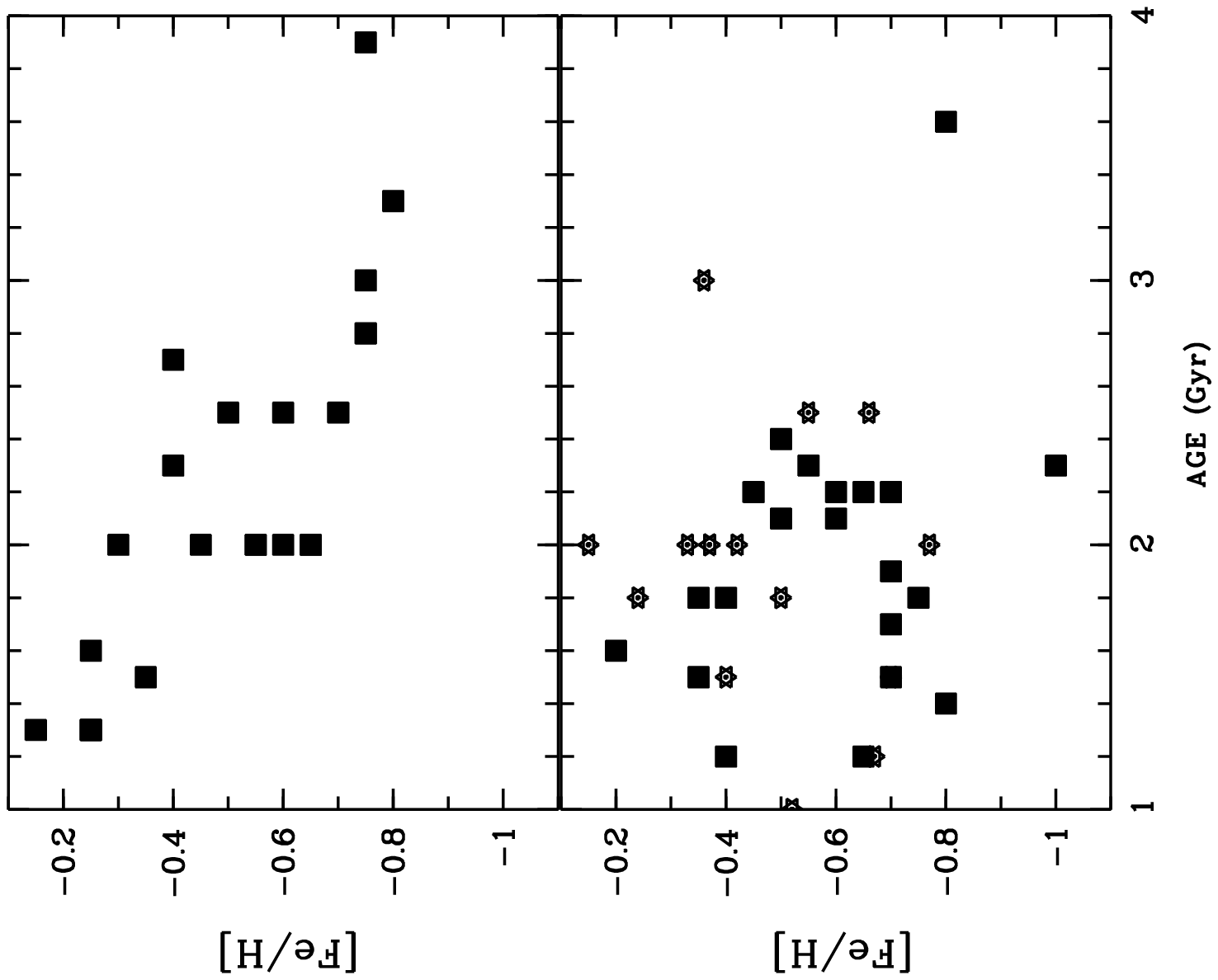}
   \caption{Upper panel: the age--metallicity relation for the LMC
intermediate--age field stars; lower panel: the same for the LMC
intermediate--age populous star clusters. The data are from \citet{piatti}
(squares) and from \citet{olszewski} (asterisks).}
         \label{agemet}
   \end{figure}

\begin{table}
\begin{center}
\caption{Data on metallicities of irregular galaxies and their MSC populations \label{tbl-1}}
\begin{tabular}{lcccccc}
\tableline\tableline
Galaxy  &M$_B$  &Gal. oxyg. abun.       & ref  & Cluster met-ty\tablenotemark{a} & ref &
Richness \\
    ~     & ~   &  12+log(O/H)      & ~           & Z  & ~ \\
\tableline
 NGC 1140  &-19.7  & 8.0--8.5& 4 & (0.008) & 4 & medium \\
 NGC 1156  &-18.0  & 8.39  & 10  &  ?    &    & medium \\
 NGC 1313  &-18.9  & 8.4   & 17   &   ?   &    & medium \\
 NGC 1569  &-17.4  & 8.17  & 12 & 0.008 & 9 & medium \\
           &       &       &     & 0.004--0.008& 1 & ~\\
 NGC 1705  &-16.1  & 8.21  & 13   & 0.008 & 2 & poor \\
 NGC 4214  &-18.4  & 8.2--8.34& 11   & 0.008 & 2 & medium \\
 NGC 4449  &-18.2  & 8.3   &  6 & 0.008 & 6 & medium \\
 NGC 5253  &-17.1  & $\sim 8.2$& 11, 14& (0.008) & 7 & poor \\
 NGC 6745  &-20.3  &  ?    &      & 0.008 & 3 & rich  \\
 LMC       &-18.1  &  8.35 &  16  & 0.007 & 5  & rich \\
 IC 10     &-16.5  &  8.22 &  16  & 0.004--0.008 &  8 & poor \\
 M 82      &-19.0  &   ?   &      &   ?   & 15 & rich \\
\tableline
\end{tabular}

\tablenotetext{a}{See text for the
comments and more detailed explanations about these values of metallicity}
\tablerefs{
(1) Anders et al. 2004; (2) Billett et al. 2002; (3) de Grijs et al. 2003; (4) de Grijs et
al. 2004; (5) Geisler et al. 2003; (6) Gelatt et al. 2001; (7) Harris et al. 2004; (8) Hunter
2001; (9) Hunter et al. 2000; (10) Hunter et al. 2001; (11) Kobulnicky \& Skillman 1996; (12)
Kobulnicky \& Skillman 1997; (13) Lee \& Skillman 2004; (14) Martin 1997; (15) Parmentier et
al. 2003; (16) Richer \& McCall 1995; (17) Walsh \& Roy 1997.}
\end{center}
\end{table}

\end{document}